\documentclass[12pt]{article}
\usepackage{dcolumn,amssymb}
\usepackage{epsfig,epsf}
\newcommand{\la}{\langle}
\newcommand{\ra}{\rangle}

\newcommand{\Z}{\mathbb{Z}}
\newcommand{\R}{\mathbb{R}}
\newcommand{\C}{\mathbb{C}}
\renewcommand{\P}{\mathbb{P}}

\newcommand{\be}{\begin{equation}}
\newcommand{\ee}{\end{equation}}
\newcommand{\bea}{\begin{eqnarray}}
\newcommand{\eea}{\end{eqnarray}}
                                                      
\newcommand{\eps}{\epsilon}
\begin{document}                                                                
\begin{flushright}
MPP-2005-18\\
\end{flushright}
\bigskip\bigskip
\begin{center}
{\Huge The clash of positivities in topological density correlators}\\
\bigskip\bigskip
{\it Miguel Aguado and Erhard Seiler\\
\bigskip
Max-Planck-Institut f\"ur Physik\\
 (Werner-Heisenberg-Institut),\\
 F\"ohringer Ring 6, 80805 Munich, Germany }
\end{center}

\begin{abstract}  
\noindent
We discuss the apparent conflict between reflection positivity and 
positivity of the topological susceptibility in two-dimensional nonlinear 
sigma models and in four-dimensional gauge theories. We pay special 
attention to the fact that this apparent conflict is already present on 
the lattice; its resolution puts some nontrivial restrictions on the 
short-distance behavior of the lattice correlator. It is found that these 
restrictions can be satisfied both in the case of asymptotic freedom and 
the dissident scenario of a critical point at finite coupling.
\end{abstract}
\vskip2mm

\section{Introduction}
Topological density correlators have some positivity 
properties that may 
seem paradoxical at first sight. If we denote the topological density by 
$q(x)$ and (minus) its two-point function (in Euclidean space) by
\be
F(x)\equiv - \la q(0) q(x)\ra,
\ee
reflection positivity (RP), i.e. positivity of the metric in Hilbert space 
demands that
\be
F(x)\ge 0 \ \ {\rm for} \ \ x\neq 0\ ,
\label{RP}
\ee
as has been pointed out long ago (see \cite{ss, s}). Actually it is easy to 
see that $F$ cannot vanish anywhere (unless it vanishes identically), i.~e. 
\be
F(x)> 0 \ \ {\rm for} \ \ x\neq 0\ ;
\label{RPs}
\ee
on account of the Lehmann-K\"all\'en spectral representation. 

On the other hand the topological susceptibility 
\be
\chi_t\equiv \int dx \la q(0) q(x)\ra = -\sum_x F(x)
\label{pos}
\ee
should be nonnegative on account of the positivity of the euclidean 
functional measure (at least if there is no nonzero $\theta$ angle), since 
it can be obtained as 
\be
\chi_t=\lim_{V\to\infty} \frac{1}{V}\la Q_V^2\ra,
\ee
where $Q_V$ is the topological charge in the finite volume $V$.
As has been stressed repeatedly, these two properties can be reconciled 
only by requiring specific contact terms in $F(x)$, something that 
is of no physical relevance in axiomatic quantum field theory, because 
contact terms do not contribute to the analytic continuation from Euclidean 
to Minkowski space.
 
We want to approach this problem by considering the quantum field theory 
as a continuum limit of a lattice field theory in which both positivities 
are already satisfied at nonzero lattice spacing. We are aware of the fact 
that (\ref{RP}) does not hold for all lattice versions of the models in 
question, but if we rely on the universality principle we should be 
allowed to restrict our attention to the lattice theories satisfying it.
After all, RP (for gauge invariant fields) has to be true in the continuum 
limit, if the theory is to make physical sense. Similarly, there are 
nonlocal definitions of the topological density that do not satisfy 
(\ref{RP}), but again the violation should only be a lattice artefact.

\section{Two dimensions}

We will discuss the case of the two-dimensional $O(3)$ nonlinear $\sigma$ 
model in some detail and remark about the $\C\P^{N-1}$ models and the 
massless and massive Schwinger models at the end. 

The lattice $O(3)$ model is defined in terms of the standard lattice 
action
\be
S=\sum_{\la xy\ra} s(x)\cdot s(y)\ ,
\ee
where $s(.)\in S^2\subset \R^3$ and the Gibbs density is proportional to
$e^{-\beta S}$. We are working on the unit lattice $\Z^2$ in a regime 
$\beta<\beta_{crt}$ where the model shows exponential clustering with 
correlation length $\xi$. The dynamically defined lattice spacing $a$ is 
proportional to the inverse correlation length
\be
a=\frac{\ell_0}{\xi}
\ee
where the constant $\ell_0$ defines the standard of length.

Of course according to the standard wisdom $\beta_{crt}=\infty$, but 
Patrascioiu and Seiler have raised doubts about this over the years (for a 
recent summary see \cite{kyoto} and references given there) and the issue 
remains an open mathematical question \cite{iamp}.  

The most natural definition of the topological density $q(x^*)$ on the 
lattice is associated with a plaquette or equivalently with a site $x^*$ of 
the dual lattice; other definitions associate $q$ with lattice sites.
As examples we mention two choices that satisfy RP:
\begin{itemize}
\item
`field theoretic definition' \cite{digiac},
\be
q_{ft}(x) = \frac{1}{32\pi} \sum_{\mu\nu}\sum_{ijk} \eps_{\mu\nu}\eps_{ijk} 
s_i(x)[s_j(x+\hat\mu)-s_j(x-\hat\mu)]\times 
[s_k(x+\hat\nu)-s_k(x-\hat\nu)]
\ee
\item
`geometric definition' \cite{bl}
\bea
q_{geom}(x^*) = && 
\frac{1}{8\pi} 
\{A(s(1),s(2),s(3))+A(s(1),s(3),s(4))\cr 
&&+A(s(1),s(2),s(4))+A(s(2),s(3),s(4))\} 
\label{geom}
\eea
where the sites 1,2,3,4 are the four corners of the plaquette dual to 
$x^*$ and $A(.,.,.)$ is the area of the spherical triangle spanned by the 
three points on the sphere appearing as arguments.
\end{itemize}
(\ref{geom}) arises from the expression found in \cite{bl} by  
symmetrization, so as to make it antisymmetric with respect to time 
reflections, a prerequisite for RP.

We study the two-point correlation function at a certain value of $\beta$, 
which we prefer to parameterize by $a(\beta)=\ell_0/\xi(\beta)$ 
\be
F_a(x)\ =\ -a^{-4} \left\la q(0)q(\frac{x}{a})\right\ra
\ee
where we inserted the prefactor $a^{-4}$ in anticipation of the continuum 
limit 
\be
F_0(x)=\lim_{a\to 0}F_a(x)\ ,
\ee
which is not expected to require any divergent field strength
renormalization.

Note that the whole lattice definition of the topological charge
density (in particular, all contact terms of the two-point correlator
arising from this definition) must be taken into account to analyze
the interplay of the behavior of the correlator at $x = 0$ and at $x
\neq 0$ necessary to fulfill positivity requirements.  For instance,
additive renormalizations suggested to define a `physical' topological
susceptibility in the continuum limit should not be introduced here.
We do not want to make any claims concerning the existence of the
continuum limit of the topological susceptibility, which is a
difficult issue in the case of the $O(3)$ model (see for instance
\cite{blatter, delia}).  The two-point correlator of the topological
charge density could be well defined in this limit even if $\chi_t$ is
not.

Since $q(x)$ is a dimension 2 operator, naively one would expect that the 
short distance behavior of its two point correlation function is
\be
F_0(x)=O\left(\frac{1}{|x|^4}\right).
\label{naive}
\ee
The two positivities satisfied by $F_a$ are 
\be
F_a(x)>0\ \ {\rm for} \ \ x\neq 0
\ee
and
\be
\chi_t^a=-\sum_x a^2 F_a(x)\geq 0\ .
\label{poschi}
\ee
These two inequalities imply 
\be
\la q(0)^2\ra\geq -\sum_{x\neq 0} \left\la q(0)q(\frac{x}{a})\right\ra
=\sum_{x\neq 0}a^2 F_a(x)\, 
\label{ineq}
\ee
and if we rewrite (\ref{poschi}) as
\be
\chi_t^a= \la q(0)^2\ra a^{-2}-\sum_{x\neq 0}a^2 F_a(x)\geq 0\ ,
\label{canc}
\ee
we see that the topological susceptibility is the remainder of the 
incomplete cancellation of the two sides of (\ref{ineq}). 

Replacing heuristically the right hand side of Eq.(\ref{ineq}) by its 
continuum limit one is tempted to write
\be
\int_{|x|\ge ad} F_0(x) d^2x \le a^{-2} \la q(0)^2\ra		
\ee
with some constant $d$ of order 1. Using the fact that according to tree 
level perturbation theory (which is uncontested) there is a constant $c$ 
such that for $\beta$ greater than some $\beta_0$
\be
\la q(0)^2\ra\leq \frac{c}{\beta^2}  
\ee
we then would conclude that 
\be
\int_{|x|\ge ad} F_0(x) d^2x \le \frac{c}{\beta^2a^2} .
\label{contineq}
\ee
We will later give a more precise derivation of a slightly weaker
inequality than Eq.(\ref{contineq}), that depends, however, on a
certain assumption about the approach to the continuum.

Note that in this equation $a$ should be considered as a function of 
$\beta$. It has to remain valid as $a\to 0$, i.e. $\beta\to\beta_{crt}$. 
So Eq.(\ref{contineq}) expresses a remarkable link between the short 
distance behavior of the topological correlator and the value of the 
critical coupling $\beta_{crt}$. If, as commonly believed, 
$\beta_{crt}=\infty$, it implies that the short distance singularity of 
$F_0(x)$ has to be softer than $1/|x|^4$. As will be discussed, this is in 
fact consistent with RG improved perturbation theory. But Eq.(\ref{contineq})
can obviously also easily be satisfied in the dissident scenario of a 
finite value of $\beta_{crt}$; in this case the `classical' behavior 
(\ref{naive}) is allowed.

Another remarkable feature in the conventional scenario is this: according 
to asymptotic scaling the topological susceptibility should be 
exponentially small in $\beta$, but the first term on the right hand side 
of (\ref{canc}) is $O(1/\beta^2)$. That means that also the second term 
has to be of that order and the cancellation between the two terms has to 
be almost complete. It has of course been known for a long time that for 
instance the geometric definition does not satisfy asymptotic scaling 
\cite{bl} numerically; it is an open question if it is satisfied for any 
definition that also obeys RP in the continuum limit. But maybe one 
should not worry about this point too much, since asymptotic scaling has 
also not been verified for the correlation length; the only interesting 
open question is the existence of a nontrivial continuum limit of 
$\chi_t^a$, which is, however, not our concern here.

Let us now turn to the derivation of (\ref{contineq}). It is certainly to 
be expected that the two-point function $F_a(x)$ converges to the continuum 
limit $F_0(x)$ pointwise. But one cannot expect that the approach is uniform
in $x$; it is to be expected that the convergence is slower the shorter the 
distance $x$ is. We make the following assumption about the approach of 
$F_a$ to the continuum: there are constants $d>0$ (independent of $a$) and 
$a_0>0$ such that  
\be
\left|\frac{F_a(x)}{F_0(x)}-1\right|\leq \frac{1}{2} \ \ {\rm for} \ a\le 
a_0\ {\rm and}\ \ell_0\ge |x|\ge a d\ .
\label{unif}
\ee 
This assumption limits the amount of nonuniformity permitted in the 
approach to the continuum; it holds for correlators of free fields 
and can be checked in perturbation theory. In principle it can also be 
tested numerically. We omitted large distances because we are considering 
the massive continuum limit and the correlation function will decay 
exponentially in $\ell_0|x|$. 
 
To use this assumption we reinterpret the lattice function $F_a(x)$ as a 
piecewise constant function in the continuum and the sum $\sum_{|x|\ge ad} 
F_a(x)$ as an integral. We get, using the triangle inequality 
\bea
&&\sum_{\ell_0\ge |x|\ge ad} F_a(x)\cr
&&\ge \int_{\ell_0\ge |x|\ge ad} F_0(x)d^2x-
\int_{\ell_0\ge|x|\ge ad}\left|F_0(x)-F_a(x)\right|d^2x \cr
&&\ge \frac{1}{2} \int_{\ell_0\ge|x|\ge ad} F_0(x)d^2x.
\eea
Inserting this in (\ref{ineq}) we get
\be
\int_{\ell_0\ge|x|\ge ad} F_0(x)d^2x \le 2a^{-2}\la q(0)^2\ra
\label{replace}
\ee
which is the announced replacement for (\ref{contineq}).

Next we discuss inequality (\ref{replace}) in the conventional 
scenario. According to RG improved tree level perturbation theory we have
(cf. \cite{bn,vicari}) 
\be
F_0(x)=g^2(x)\frac{1}{|x|^4}+O(g^3(x))\ \ \ {\rm for} \ x\to 0. 
\label{rgpt}
\ee
Inserting the leading order perturbative running coupling
\be
g^2(x)\sim\frac{\rm const}{(\ln\mu|x|)^2} 
\ee
we get
\be
F_0(x)\sim\frac{\rm const}{|x|^4 (\ln\mu|x|)^2}\  ,
\ee
i.e. the short distance behavior is indeed softer than the naive one.
It is now not hard to see that with this behavior one gets
\be
\int_{\ell_0\ge|x|\ge ad} F_0(x)d^2x= 
O\left(\frac {a^{-2}}{(\ln (\mu ad))^2}\right). 
\ee
This is consistent with (\ref{replace}) if one assumes asymptotic 
scaling, because then to leading order $\beta^2=O((\ln a)^2)$.

The above discussion carries over without any essential changes to the 
two-dimensional $\C\P^{N-1}$ models; in fact it is even simpler due to the 
fact that there is a very natural definition of the topological density 
as the field strength of the auxiliary abelian gauge field in these models.

In the (massive or massless) Schwinger model the situation is slightly 
different: the value of $\beta_{crt}$ is finite; in the massless version 
there is perfect cancellation between the two terms in (\ref{replace}), 
whereas in the massive Schwinger model the cancellation is incomplete. 
The Schwinger model is also atypical in that the topological density is 
really a dimension 0 field -- this is due to the fact that there is a 
dimensional parameter (the electric charge) in this model. 

\section{Four dimensions}

The discussion in four dimensions, in particular QCD, parallels the 
one in two dimensions, so we will limit ourselves to pointing out the 
necessary modifications of the previous discussion.

Again there are different lattice definitions of the topological density to 
be considered. Among them the so-called field theoretic definition \cite{div}
satisfies RP in a straight-forward manner. There are also geometric 
definitions satisfying RP \cite{l,lasher}. The physically most relevant 
definitions, however, are based on the relation between chirality and 
topology; only these lead to a solution of the $U(1)$ problem of QCD via 
credible derivations of the Witten-Veneziano formula 
\cite{w,v,giust1,giust2, luscher, deldebbio,s}, 
and are generally nonlocal, making RP very nonobvious. In this context it 
is gratifying that recently the topological two point function based on 
the overlap Dirac operator has been measured and found indeed to satisfy 
RP, at least for lattice distances greater than 2 \cite{horvath}. 

The topological density, being given by 
$\frac {g^2}{32}\pi^2 F_{\mu\nu}\tilde F_{\mu\nu}$ in the continuum, is 
now a dimension 4 operator and hence its two point correlator on the 
lattice should be defined as
\be
-F_a(x) = a^{-8} \la q(0)q(\frac{x}{a})\ra,
\ee
where $x$ may be a site of the original or the dual lattice. The short 
distance behavior of the continuum limit $F_0(x)$ is now naively
\be
F_0(x)=O\left(\frac{1}{|x|^8}\right)
\ee
and the topological susceptibility is the difference of two almost 
cancelling positive terms:
\be
\chi_t^a=\la q(0)^2\ra a^{-4}-\sum_{x\neq 0}a^4  F_a(x)\ 
\ee
as in two dimensions. Again the contact term satisfies
\be
\la q(0)^2\ra\leq \frac{c}{\beta^2} 
\ee
just as in two dimensions. 

The approach to the continuum should satisfy the same uniformity as in two 
dimensions (see Eq.(\ref{unif})). By the same reasoning as above we obtain 
then
\be
\int_{\ell_0\ge|x|\ge ad} F_0(x)d^4x \le 2a^{-4}\la q(0)^2\ra\ 
\ee 
and again we find that this can be satisfied either by assuming the softened 
short distance behavior 
\be
F_0(x)\sim \frac{1}{|x|^8(\ln(\mu|x|))^2}
\ee
or, of course, by the existence of a critical point at finite $\beta$.

\section{Conclusions}

The two positivities of the topological two-point function are superficially 
in conflict with each other. To reconcile them, one needs first of all 
specific contact terms. It is a remarkable fact that we obtain restrictions 
on non-universal `unphysical' quantities from these considerations.

In addition we found out that:
\begin{itemize}
\item either the short distance behavior of $F_0(x)$ is softened 
logarithmically compared to the naive tree level behavior,  in a way 
consistent with RG improved tree level perturbation theory,
\item or there is a critical point at a finite value of $\beta$.
\end{itemize}
In another paper \cite{as} we report on a direct lattice 
perturbation calculation for the $2D$ $O(3)$ model, which verifies 
consistency with the RG improved tree level expression Eq. (\ref{rgpt}).

{\it Acknowledgement:} We thank Peter Weisz for a useful discussion.

\end{document}